\documentclass[useAMS,usenatbib]{mn2e}

\usepackage{latexsym,graphicx,natbib}

\usepackage{color}

\newcommand\kms{{\rm\,km\,s^{-1}}}
\newcommand\msun{\rm\,M_\odot}

\newcommand\lsun{\rm\,L_\odot}
\newcommand\hii{H\,{\sc ii} \,}

\DeclareRobustCommand{\ion}[2]{%
\relax\ifmmode
\ifx\testbx\f
{\mathrm{#1\,\textsc{#2}}}\else {\mathrm{#1\,\mathsc{#2}}}\fi
\else\textup{#1\,{\mdseries\textsc{#2}}}%
\fi}
\newcommand{\MC}{\multicolumn}

\def\apgt{\ {\raise-.5ex\hbox{$\buildrel>\over\sim$}}\ }
\def\aplt{\ {\raise-.5ex\hbox{$\buildrel<\over\sim$}}\ }

\title[New bona fide luminous blue variable in Norma]{Discovery of a new bona fide luminous blue variable in Norma\footnotemark[0]\thanks{Based on observations
made with the Southern African Large Telescope (SALT) under the
program \mbox{2015-1-SCI-017} (PI: A.Y.~Kniazev).}}
\author[V.V.Gvaramadze et al.]
       {V. V.~Gvaramadze,$^{1,2,3}$\thanks{E-mail: vgvaram@mx.iki.rssi.ru (VVG)},
        A. Y.~Kniazev,$^{4,5,1}$ and L. N.~Berdnikov$^{1,6,3}$ \\
        $^{1}$Sternberg Astronomical Institute, Lomonosov Moscow State University, Universitetskij Pr. 13, Moscow 119992, Russia\\
        $^{2}$Space Research Institute, Russian Academy of Sciences, Profsoyuznaya 84/32, 117997 Moscow, Russia \\
        $^{3}$Isaac Newton Institute of Chile, Moscow Branch, Universitetskij Pr. 13, Moscow 119992, Russia \\
        $^{4}$South African Astronomical Observatory, PO Box 9, 7935 Observatory, Cape Town, South Africa \\
        $^{5}$Southern African Large Telescope Foundation, PO Box 9, 7935 Observatory, Cape Town, South Africa \\
        $^{6}$Astronomy and Astrophysics Research division,  Entoto Observatory and Research Center, P.O.Box 8412, Addis Ababa, Ethiopia \\
        }
\begin{document}

\date{Accepted 2015 September 29.  Received 2015 September 29; in original form 2015 August 11}


\maketitle

\label{firstpage}

\begin{abstract}
We report the results of optical spectroscopy of the candidate
evolved massive star MN44 revealed via detection of a circular
shell with the {\it Spitzer Space Telescope}. First spectra taken
in 2009 May--June showed the Balmer lines in emission as well as
numerous emission lines of iron, which is typical of luminous blue
variables (LBVs) near the visual maximum. New observations carried
out in 2015 May--September detected significant changes in the
spectrum, indicating that the star became hotter. We found that
these changes are accompanied by significant brightness
variability of MN44. In particular, the $I_{\rm c}$-band
brightness decreased by $\approx$ 1.6 mag during the last six
years and after reaching its minimum in 2015 June has started to
increase. Using archival data, we also found that the $I_{\rm
c}$-band brightness increased by $\approx$3 mag in $\approx$30 yr
preceding our observations. MN44 therefore represents the
seventeenth known example of the Galactic bona fide LBVs. We
detected a nitrogen-rich knot to the northwest of the star, which
might represent an interstellar cloudlet interacting with the
circumstellar shell. We discuss a possible association between
MN44 and the {\it INTEGRAL} transient source of hard X-ray
emission IGR\,J16327$-$4940, implying that MN44 might be either a
colliding-wind binary or a high-mass X-ray binary.
\end{abstract}


\section{Introduction}
\label{sec:intro}

\begin{figure*}
\begin{center}
\includegraphics[width=14cm,angle=0]{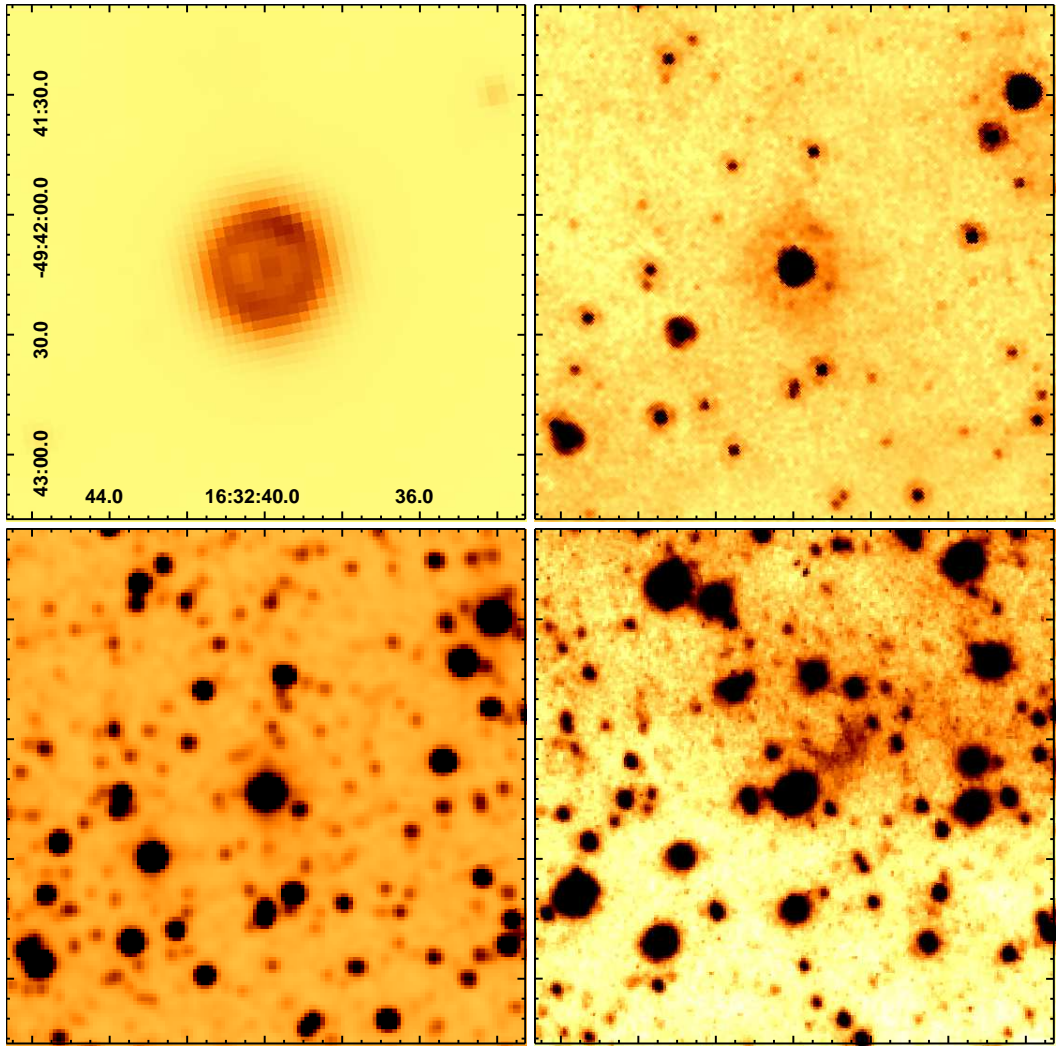}
\end{center}
\caption{From left to right, and from top to bottom: {\it Spitzer}
MIPS $24 \, \mu$m and IRAC 8\,$\mu$m, 2MASS $K_{\rm s}$-band and
SHS H$\alpha$+[N\,{\sc ii}] images of the region containing the
circular shell MN44 and its central star (the scale of the images
is the same). The coordinates are in units of RA (J2000) and Dec.
(J2000) on the horizontal and vertical scales, respectively.
    }
\label{fig:neb}
\end{figure*}

Some massive stars evolve through the so-called luminous blue
variable (LBV) phase in the course of their life (Conti 1984).
During this phase, they experience episodes of enhanced (sometimes
eruptive) mass loss, which is manifested in appearance of strong
emission lines in their spectra. These episodes, lasting from
years to decades and more, are accompanied by major brightness and
spectral variability -- the defining characteristics of the very
rare class of objects called LBV stars (Humphreys \& Davidson
1994). Recent searches for new members of this class though
detection of their circumstellar nebulae (a common attribute of
the LBVs; Clark, Larionov \& Arkharov 2005; Kniazev, Gvaramadze \&
Berdnikov 2015) resulted in discovery of several dozens of compact
mid-infrared nebulae of various morphology (Gvaramadze, Kniazev \&
Fabrika 2010a), whose central stars exhibit rich emission spectra,
typical of bona fide LBVs (Gvaramadze et al. 2010a,b, 2012a;
Wachter et al. 2010, 2011; Stringfellow et al. 2012a,b; Flagey et
al. 2014). These discoveries nearly doubled the known population
of Galactic candidate LBVs (cLBVs). To prove their LBV status a
significant spectral and photometric variability has to be
detected. Disclosure of new bona fide LBVs is of high importance
for understanding the nature of their variability and the role of
the LBV phase in the stellar evolution.

In this paper, we present recent results of our ongoing
spectroscopic and photometric monitoring of the newly identified
cLBVs, aimed to detect changes in their spectra and brightness.
Our monitoring campaign, started in 2009, has already resulted in
discovery of two new Galactic bona fide LBVs -- Wray\,16-137
(Gvaramadze et al. 2014) and WS1 (Kniazev et al. 2015). Here, we
report the discovery of the third Galactic bona fide LBV -- MN44,
whose LBV candidacy was suggested by the presence of a circular
shell around it (Gvaramadze et al. 2010a). In
Section\,\ref{sec:neb}, we present for the first time the images
of the shell and its central star at several wavelengths, and
review the existing data on the central star. In
Section\,\ref{sec:obs} we describe our spectroscopic and
photometric observations. The results and implications are
discussed in Section\,\ref{sec:dis}. We summarize and conclude in
Section\,\ref{sec:con}.

\begin{table*}
  \caption{Journal of the observations.}
  \label{tab:log}
  \renewcommand{\footnoterule}{}
  \begin{tabular}{llccccccccc}
      \hline
Spectrograph & Date & Exposure & Spectral scale & Spatial scale & Slit/Seeing & Spectral range \\
& & (min) & (\AA \, pixel$^{-1}$) & (arcsec pixel$^{-1}$) & (arcsec) & (\AA) \\
      \hline
Cassegrain (SAAO 1.9-m)&2009 May 31      & 3$\times$15 & 2.3 & 1.4 & 2.0$\times$180/1.0 & 4200$-$8100\\
Cassegrain (SAAO 1.9-m)&2009 June 2      & 3$\times$20 & 2.3 & 1.4 & 2.0$\times$180/1.3 & 4200$-$8100\\
RSS (SALT)             &2015 May 6       & 1.5+15      & 0.97& 0.51& 1.25$\times$480/1.0 & 4200$-$7300\\
RSS (SALT)             &2015 June 14     & 1+20        & 0.97& 0.25& 1.25$\times$480/2.5 & 4350$-$7450\\
RSS (SALT)             &2015 August 2    & 1+25        & 0.97& 0.25& 1.25$\times$480/2.0 & 4200$-$7300\\
RSS (SALT)             &2015 September 8 & 1+25        & 0.97& 0.25& 1.25$\times$480/1.4 & 4200$-$7300\\
      \hline
    \end{tabular}
    \end{table*}

\section{Circular shell MN44 and its central star}
\label{sec:neb}

The circular shell MN44\footnote{In the Set of Identifications,
Measurements and Bibliography for Astronomical Data (SIMBAD) data
base this shell is named [GKF2010]\,MN44.} is located in the Norma
constellation and it is one of many dozens of mid-infrared compact
nebulae discovered by Gvaramadze et al. (2010a) in the $24 \,
\mu$m archival data of the {\it Spitzer Space Telescope} obtained
with the Multiband Imaging Photometer for {\it Spitzer} (MIPS;
Rieke et al. 2004) within the framework of the 24 and 70 Micron
Survey of the Inner Galactic Disk with MIPS (Carey et al. 2009).
In the MIPS 24\,$\mu$m image MN44 appears as a limb-brightened
circular shell of radius and thickness of $\approx$15 and 6
arcsec, respectively, centred on a point-like source (see
Fig.\,\ref{fig:neb}). The central point source (star) is also
visible in all (3.6, 4.5, 5.8 and $8.0\,\mu$m) images obtained
with the {\it Spitzer} Infrared Array Camera (IRAC; Fazio et al.
2004) within the Galactic Legacy Infrared Mid-Plane Survey
Extraordinaire (Benjamin et al. 2003), as well as in all
($J,H,K_{\rm s}$) Two-Micron All Sky Survey (2MASS) images
(Skrutskie et al., 2006). The 2MASS coordinates of this star are:
RA(J2000)=$16^{\rm h} 32^{\rm m} 39\fs95$, Dec.(J2000)=$-$$49\degr
42\arcmin 13\farcs8$ and $l$=$335\fdg0635$, $b$=$-$$1\fdg1557$. In
the following, we will use the name MN44 only for the central
star.

Fig.\,\ref{fig:neb} shows that the nebula could also be discerned
in the IRAC 8\,$\mu$m image and that there is the gleam of
emission around the star in the H$\alpha$+[N\,{\sc ii}] image
obtained in the framework of the SuperCOSMOS H-alpha Survey (SHS;
Parker et al. 2005). The SHS image also shows a knot of enhanced
emission to the northwest of the star. This knot partly delineates
the diffuse emission around the star and then curves in the
northeast direction. Comparison of the 24 and 8\,$\mu$m images
with the SHS one shows that the shell is brighter at the place of
possible contact with the knot, which might be caused by
interaction of the shell with a density inhomogeneity in the
ambient medium. The nature of the optical emission around MN44 is
further discussed in Section\,\ref{sec:ha}.

MN44 was identified as an emission-line star by Vega et al.
(1980), who searched for H$\alpha$-emission objects using
objective-prism spectra. In the SIMBAD data base this star is
named EM*\,VRMF\,55. According to Vega et al. (1980), in 1974 the
$V$-band magnitude of MN44 was fainter than $15.58$. MN44 is
located within the error circle of radius of 4.5 arcmin of the
{\it INTEGRAL} transient source of hard X-ray emission
IGR\,J16327$-$4940 (Bird et al. 2010). Masetti et al. (2010)
obtained an optical spectrum of MN44 on 2009 August 10 using the
1.9-m telescope of the South African Astronomical Observatory
(SAAO) (see their fig.\,7) and broadly classified it as an OB star
due to the strength of its H$\alpha$ emission line, ``which is
much larger than the typical values seen in blue supergiants".
These authors tentatively associated MN44 with IGR\,J16327$-$4940
and classified this X-ray source as a high-mass X-ray binary
(HMXB) because of its ``overall early-type star spectral
appearance, which is typical of this class of objects". Possible
implications of this association are discussed in
Section\,\ref{sec:bin}.

\begin{figure*}
\begin{center}
\includegraphics[width=9cm,angle=270,clip=]{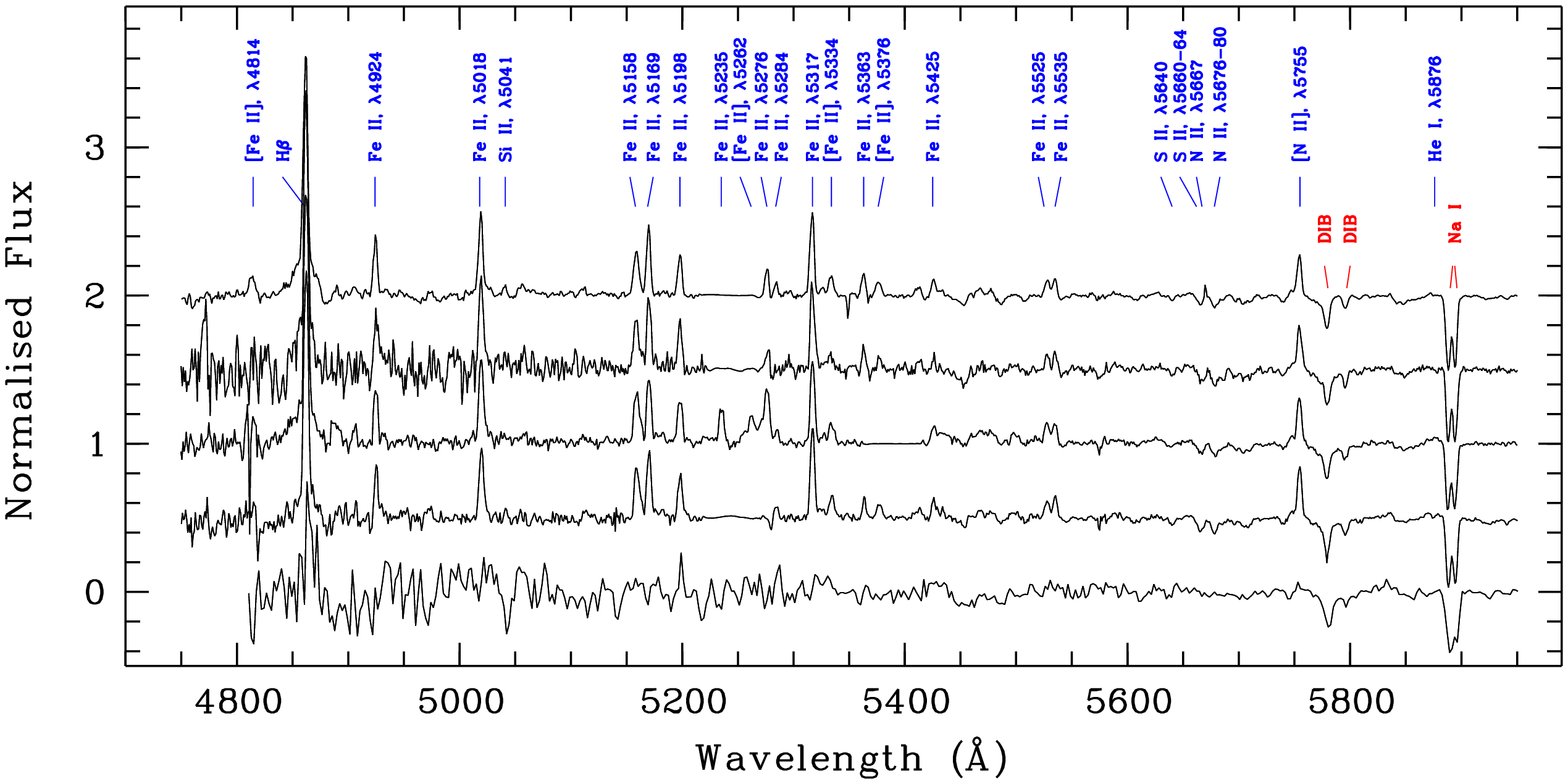}
\includegraphics[width=9cm,angle=270,clip=]{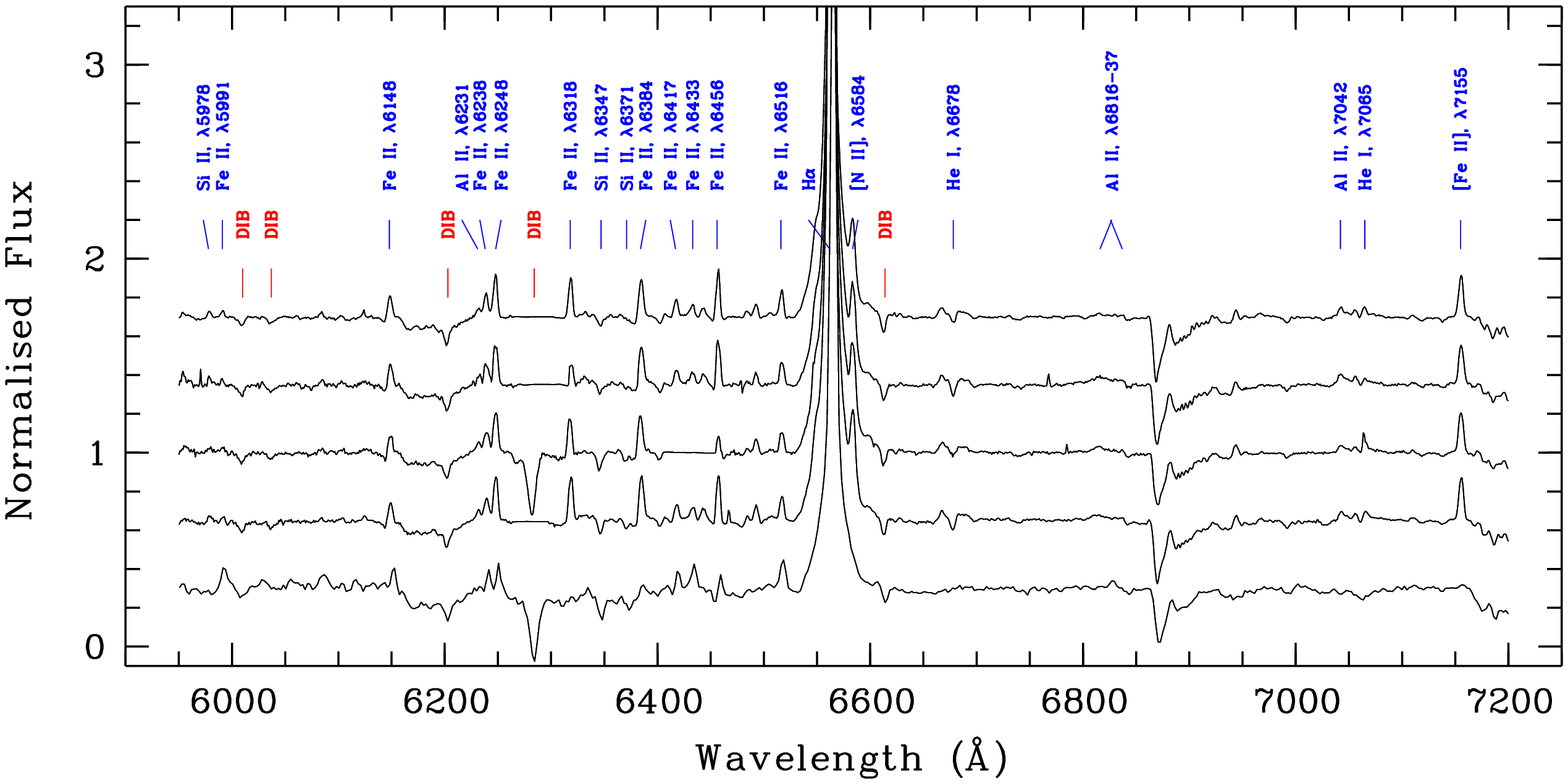}
\end{center}
\caption{Normalized spectra of MN44 obtained in 2009 with the SAAO
1.9-m telescope (the lower curve) and on 2015 May 6, June 14,
August 2 and September 8 with the SALT (four upper curves with the
date increasing from bottom to top). The principal lines and most
prominent DIBs are indicated.} \label{fig:spec}
\end{figure*}

\section{Observations}
\label{sec:obs}

First spectra of MN44 were obtained with the SAAO 1.9-m telescope
on 2009 May 31 and June 2. The observations were performed with
the Cassegrain spectrograph using a slit of 3 arcmin $\times$ 2
arcsec and grating with 300 lines mm$^{-1}$. This spectral setup
covered a wavelength range of $\approx$4200--8100\,\AA\ with a
reciprocal dispersion of $\approx$2.3\,\AA\,pixel$^{-1}$ and the
spectral resolution full width at half-maximum (FWHM) of
$\approx$7\,\AA. The slit was oriented at the parallactic angle
because there was no possibility to select a position angle (PA)
at this telescope. Three exposures of 900\,s and 1200 s were taken
during the first and second nights, respectively, with a seeing of
$\approx$1.0--1.3 arcsec. The data were reduced in a standard way
using the {\sc midas} and {\sc iraf} software. All one-dimensional
(1D) spectra obtained during the two nights were then averaged to
produce a final spectrum (see the lower curve in
Fig.\,\ref{fig:spec}). The spectrum shows the H$\alpha$ and
H$\beta$ emission lines and numerous singly ionized iron emission
lines, typical of LBVs near the brightness maximum (e.g. Stahl et
al. 2001). In the red part of the spectrum we detected the
prominent O\,{\sc i} $\lambda$7771--4 \AA \, triplet in absorption
(not shown in Fig.\,\ref{fig:spec}), which is a good indicator of
luminosity (Merrill 1934). The equivalent width (EW) of this
triplet of 1.93$\pm$0.18 \AA \, implies that MN44 is a supergiant
star of luminosity class Ia (e.g. Keenan \& Hynek 1950; Osmer
1972). These findings along with the presence of the circular
shell around the star allowed us to classify MN44 as a cLBV.

To search for possible spectral variability of MN44, we obtained
four more spectra with the Southern African Large Telescope (SALT;
Buckley, Swart \& Meiring 2006; O'Donoghue et al. 2006) on 2015
May 6, June 14, August 2 and September 8 (see Table\,\ref{tab:log}
for the log of our spectroscopic observations of MN44). The
spectra were taken with the Robert Stobie Spectrograph (RSS; Burgh
et al. 2003; Kobulnicky et al. 2003) in the long-slit mode with a
$1.25\arcsec$ slit width. The PG900 grating was used on May 6,
August 2 and September 8 to cover the spectral range of 4200--7300
\AA. We used the same grating on June 14 to cover the spectral
range of 4350--7450 \AA, which allowed us to fill gaps in the
wavelength coverage of the three other spectra. The final
reciprocal dispersion and FWHM spectral resolution of the spectra
are $0.97$ \AA \, pixel$^{-1}$ and 4.51$\pm$0.13~\AA,
respectively. The RSS uses a mosaic of three 2048$\times$4096 CCDs
and the final spatial scales for observations were
0.51\arcsec\,pix$^{-1}$ in the first night and
0.25\arcsec\,pix$^{-1}$ in the next three nights. The
corresponding seeing was of $\approx$1.0, 2.5, 2.0 and 1.4 arcsec.
Two exposures were taken each night. The short exposures were used
to avoid saturation of the extremely strong H$\alpha$ emission
line. In all observations the slit was oriented at PA=43$\degr$
(measured from north to west) in order to cross the region of
bright emission (knot) to the northwest of MN44. A Xe lamp arc
spectrum was taken immediately after the science frames. A
spectrophotometric standard star was observed during twilight time
for relative flux calibration. The primary reduction of the data
was done with the SALT science pipeline (Crawford et al. 2010).
After that, the bias and gain corrected and mosaiced long-slit
data were reduced in the way described in Kniazev et al. (2008).
The resulting normalized spectra are presented in
Fig.\,\ref{fig:spec} (see the upper four curves). In the figure we
show only parts of the spectra longward of $\approx$4800\, \AA \,
because most of the SALT spectra were taken during bright time and
their blue parts are very noisy and do not contain meaningful
information.

\begin{table*}
  \caption{Photometry of MN44.}
  \label{tab:phot}
  \renewcommand{\footnoterule}{}
  \begin{center}
  \begin{minipage}{\textwidth}
  \begin{tabular}{lcccccc}
      \hline
      Date & $B$ & $V$ & $I_{\rm c}$ & $V-I_{\rm c}$ & $J$ & $K_{\rm s}$ \\
      \hline
      1974 May$^{(1)}$ & -- & $>15.58$ & -- &  -- & -- & -- \\
      1980 May 23$^{(2)}$ & -- & -- & 13.3$\pm$0.2 & -- & -- & -- \\
      1982 August 8$^{(2)}$ & 18.9$\pm$0.3 & -- & -- & -- & -- & -- \\
      1998 July 1$^{(3)}$ & -- & -- & 11.84$\pm$0.03 & -- & 9.01$\pm$0.07 & 7.42$\pm$0.06 \\
      1998 November 3$^{(3)}$ & -- & -- & 11.74$\pm$0.03 & -- & 8.90$\pm$0.08 & 7.28$\pm$0.06 \\
      1999 June 18$^{(4)}$ & -- & -- & -- & -- & 8.41$\pm$0.02 & 6.81$\pm$0.03 \\
      2009 May 31$^{(5)}$ & -- & 14.41$\pm$0.04 & 10.20$\pm$0.10 & 4.21$\pm$0.11 & -- & -- \\
      2012 May 5$^{(6)}$ & -- & 14.75$\pm$0.01 & 10.46$\pm$0.01 & 4.29$\pm$0.01 & -- & -- \\
      2013 January 13$^{(6)}$ & 17.86$\pm$0.20 & 14.99$\pm$0.01 & 10.74$\pm$0.01 & 4.25$\pm$0.01 & -- & -- \\
      2014 April 25$^{(6)}$ & 18.58$\pm$0.08 & 15.67$\pm$0.03 & 11.47$\pm$0.01 & 4.20$\pm$0.03 & -- & -- \\
      2015 May 19$^{(7)}$ & -- & 15.70$\pm$0.03 & 11.77$\pm$0.05 & 3.93$\pm$0.06 & -- & -- \\
      2015 June 14$^{(8)}$ & -- & -- & 11.81$\pm$0.06 & -- & -- \\
      2015 August 2$^{(8)}$ & -- & -- & 11.70$\pm$0.05 & -- & -- \\
      2015 September 8$^{(8)} $ & -- & -- & 11.65$\pm$0.04 & -- & -- \\
      \hline
    \end{tabular}
    \end{minipage}
    \end{center}
     (1) Vega et al. (1980); (2) USNO B-1; (3) DENIS; (4) 2MASS; (5) 1.9-m telescope; (6) 76-cm telescope;
     (7) 1-m telescope; (8) SALT.
    \end{table*}

To detect photometric variability of MN44, we determined its $B,
V$ and $I_{\rm c}$ magnitudes on CCD frames obtained with the SAAO
76-cm telescope in 2012$-$2014. We used an SBIG ST-10XME CCD
camera equipped with $BVI_{\rm c}$ filters of the Kron-Cousins
system (see e.g. Berdnikov et al. 2012) to build the system of
secondary standards in the field of MN44. With these standards, we
derived $V$ and $I_{\rm c}$ magnitudes of MN44 using images
obtained with the SAAO 1-m telescope on 2015 May 19, and an
$I_{\rm c}$ magnitude using acquisition images obtained with the
SALT on 2015 June 14, August 2 and September 8. We also calibrated
the 1.9-m telescope spectrum and synthesized its $V$ magnitude in
the way described in Kniazev et al. (2005). Using the same
standards, we recalibrated the $I$ and $B$ magnitudes from the
USNO\,B-1 catalogue (Monet et al. 2003). The results are presented
in Table\,\ref{tab:phot}. To this table we also added the lower
limit on the $V$ magnitude given in Vega et al. (1980), $J$ and
$K_{\rm s}$ magnitudes from 2MASS and two-epoch $I, J$ and $K_{\rm
s}$ photometry from the Deep Near Infrared Survey of the Southern
Sky (DENIS; The DENIS Consortium, 2005).

\section{Discussion}
\label{sec:dis}

\subsection{MN44: a bona fide LBV}
\label{sec:lbv}

Fig.\,\ref{fig:spec} shows a montage of five spectra obtained in
2009 and 2015 with the principal lines and most prominent diffuse
interstellar bands (DIBs) indicated. All wavelengths are given in
air. The lower curve represents the averaged spectrum based on two
observations carried out on 2009 May 31 and June 2 with the SAAO
1.9-m telescope. Four other curves (from bottom to top) correspond
to spectra taken with the SALT on 2015 May 6, June 14, August 2
and September 8. EWs, FWHMs and heliocentric radial velocities
(RVs) of some lines in the spectra (measured applying the {\sc
midas} programs; see Kniazev et al. 2004 for details) are given in
Table\,\ref{tab:inten}.

\begin{table}
\centering{ \caption{EWs, FWHMs and RVs of some lines in the
spectra of MN44.} \label{tab:inten}
\begin{tabular}{lrcr}
\hline \rule{0pt}{10pt}
$\lambda_{0}$(\AA) Ion & \MC{1}{c}{EW($\lambda$)} & FWHM($\lambda$) & \MC{1}{c}{RV} \\
& \MC{1}{c}{(\AA)} & \MC{1}{c}{(\AA)} & \MC{1}{c}{($\kms$)} \\
\hline  \rule{0pt}{10pt} & \MC{2}{c}{2009 May 31 -- June 2 (averaged)}     \\
%
4861\ H$\beta$\        &  1.46$\pm$0.50  &  5.41$\pm$2.12 &   206$\pm$58 \\
6347\ Si\ {\sc ii}\    &$-$0.25$\pm$0.15 &   ---          &    35$\pm$15 \\
6371\ Si\ {\sc ii}\    &$-$0.19$\pm$0.15 &   ---          &      ---     \\
6563\ H$\alpha$\       & 38.41$\pm$1.70  &  8.05$\pm$0.41 &   134$\pm$16 \\
\hline \rule{0pt}{10pt} & \MC{2}{c}{2015 May 6}     \\ 
4861\ H$\beta$\        &  9.24$\pm$0.26  &  5.18$\pm$0.16 &   60$\pm$15  \\
5334\ [Fe\ {\sc ii}]\  &  0.61$\pm$0.03  &  5.74$\pm$0.31 &   53$\pm$15  \\
5376\ [Fe\ {\sc ii}]\  &  0.53$\pm$0.03  &  5.39$\pm$0.26 &   50$\pm$15  \\
5755\ [N\ {\sc ii}]\   &  1.68$\pm$0.09  &  4.48$\pm$0.27 &    9$\pm$14  \\
6347\ Si\ {\sc ii}\    &$-$0.29$\pm$0.03 &  4.17$\pm$0.29 &$-$26$\pm$11  \\
6371\ Si\ {\sc ii}\    &$-$0.08$\pm$0.02 &  4.78$\pm$0.40 & $-$8$\pm$11  \\
6563\ H$\alpha$\       &112.01$\pm$1.62  &  5.15$\pm$0.09 &   57$\pm$11  \\
7155\ [Fe\ {\sc ii}]\  &  1.15$\pm$0.03  &  5.04$\pm$0.09 &   32$\pm$10  \\
\hline \rule{0pt}{10pt} & \MC{2}{c}{2015 June 14}     \\ 
4861\ H$\beta$\        & 12.62$\pm$0.50  &  5.53$\pm$0.25 &   53$\pm$16 \\
5334\ [Fe\ {\sc ii}]\  &  0.54$\pm$0.08  &  6.21$\pm$0.62 &   16$\pm$20 \\
5755\ [N\ {\sc ii}]\   &  1.59$\pm$0.11  &  4.81$\pm$0.27 &$-$21$\pm$14 \\
6347\ Si\ {\sc ii}\    &$-$0.49$\pm$0.08 &  5.62$\pm$0.22 &$-$83$\pm$12 \\
6371\ Si\ {\sc ii}\    &$-$0.08$\pm$0.06 &  4.76$\pm$0.60 &$-$68$\pm$12 \\
6563\ H$\alpha$\       &117.92$\pm$1.92  &  5.31$\pm$0.10 &   43$\pm$11 \\
7155\ [Fe\ {\sc ii}]\  &  1.18$\pm$0.08  &  5.19$\pm$0.10 &    6$\pm$10 \\
\hline \rule{0pt}{10pt} & \MC{2}{c}{2015 August 2}     \\ 
4861\ H$\beta$\        &  9.81$\pm$0.50  &  5.14$\pm$0.30 &   23$\pm$11 \\
5334\ [Fe\ {\sc ii}]\  &  0.23$\pm$0.09  &  6.51$\pm$1.18 & $-$67$\pm$26 \\
5376\ [Fe\ {\sc ii}]\  &  0.32$\pm$0.09  &  3.80$\pm$0.61 &   17$\pm$16 \\
5755\ [N\ {\sc ii}]\   &  1.46$\pm$0.13  &  5.04$\pm$0.26 & $-$31$\pm$8 \\
6347\ Si\ {\sc ii}\    &$-$0.03$\pm$0.07 &  3.17$\pm$0.23 & $-$87$\pm$7 \\
6371\ Si\ {\sc ii}\    &$-$0.02$\pm$0.07 &  3.07$\pm$0.63 & $-$65$\pm$7 \\
6563\ H$\alpha$\       &104.11$\pm$1.84  &  5.27$\pm$0.11 &   27$\pm$6  \\
7155\ [Fe\ {\sc ii}]\  &  1.08$\pm$0.11  &  5.38$\pm$0.11 & $-$2$\pm$5  \\
\hline \rule{0pt}{10pt} & \MC{2}{c}{2015 September 8}     \\ 
4815\ [Fe\ {\sc ii}]\     &  0.78$\pm$0.07 &  5.98$\pm$0.59  & $-$38$\pm$20 \\
4861\ H$\beta$\           & 12.16$\pm$0.47 &  5.60$\pm$0.25  & $-$14$\pm$14 \\
5334\ [Fe\ {\sc ii}]\     &  0.49$\pm$0.02 &  5.28$\pm$0.22  & $-$24$\pm$12 \\
5376\ [Fe\ {\sc ii}]\     &  0.45$\pm$0.03 &  5.48$\pm$0.32  & $-$37$\pm$13 \\
5755\ [N\ {\sc ii}]\      &  1.24$\pm$0.05 &  5.05$\pm$0.20  & $-$45$\pm$11 \\
6347\ Si\ {\sc ii}\       &--0.10$\pm$0.05 &  4.60$\pm$0.34  & $-$66$\pm$9  \\
6563\ H$\alpha$\          &102.16$\pm$1.92 &  4.71$\pm$0.10  &  $-$4$\pm$11 \\
7155\ [Fe\ {\sc ii}]\     &  1.16$\pm$0.02 &  4.92$\pm$0.09  & $-$16$\pm$9  \\
\hline \MC{4}{p{8cm}}{{\it Note}: The negative EWs correspond to
absorption lines.}
\end{tabular}
 }
\end{table}

Comparison of the spectra shows that in 2015 the EWs of the
H$\alpha$ and H$\beta$ lines increased by $\approx$3 and 6--9
times, respectively, compared to 2009 (see also
Table\,\ref{tab:inten}), while the broad Thomson scattering wings
of these lines became even more broader. Some Fe\,{\sc ii} lines
already present in the 2009's spectrum, e.g. at
$\lambda\lambda$5317, 6384, 6456, became very prominent in 2015.
The forbidden lines of [N\,{\sc ii}] $\lambda$5755 and [Fe\,{\sc
ii}] $\lambda\lambda$5334, 5376 and 7155 increased their strength
as well, and the [N\,{\sc ii}] $\lambda$6584 line became very
prominent on the red wing of the H$\alpha$ line. More importantly,
the 2015's spectra show emergence of the He\,{\sc i} lines
$\lambda\lambda$5876, 6676 and 7065. These lines were absent in
the first epoch spectrum, but show up in 2015 May 6 as central
absorptions accompanied by blue and red emission wings. The
presence of these lines imply that the star became hotter during
the last six years, which is expected in view of the significant
decline of the stellar brightness during the same time period (see
below). Fig.\,\ref{fig:HeI} plots the evolution of the He\,{\sc i}
$\lambda$6676 line profile with time. One can see that the cental
absorption became less deep in June 14, then returned to almost
the same state as it was in May 6 and then again became less deep,
while the emission wings remain almost intact.

\begin{figure}
\begin{center}
\includegraphics[angle=270,width=1.0\columnwidth,clip=]{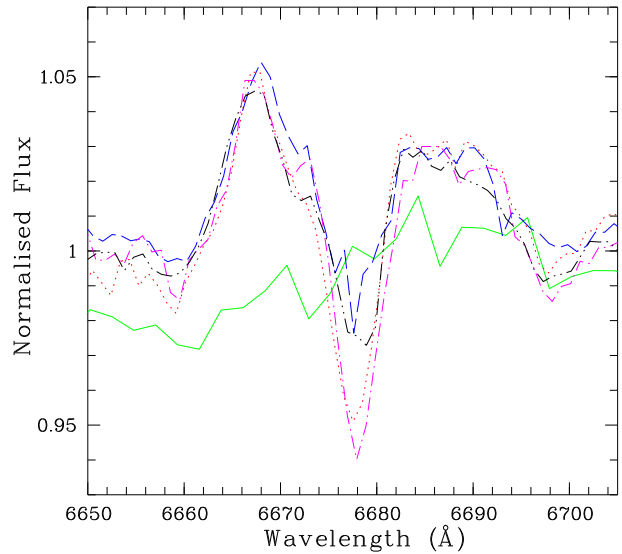}
\end{center}
\caption{Evolution of the He\,{\sc i} $\lambda$6678 line profile
with time: 2009 May 31 -- June 2 (solid green line), 2015 May 6
(dotted red line), 2015 June 14 (dashed blue line), 2015 August 2
(dash-dotted violet line) and 2015 September 8 (dash-dot-dotted
black line).} \label{fig:HeI}
\end{figure}

The [Fe\,{\sc ii}] $\lambda$5376 line in the spectra taken on 2015
May 6 and September 8 has a distinct flat-topped profile. This
indicates that the line is formed in a region of constant
expansion velocity and therefore its FWHM (respectively,
5.39$\pm$0.26 \AA \, and 5.48$\pm$0.32 \AA) could be used to
estimate the terminal wind velocity, $v_\infty$ (Stahl et al.
1991). After correction for instrumental width, one finds
$v_\infty$=165$\pm$29 and 174$\pm$33 $\kms$. The wind velocity
could also be estimated by using the FWHM of the [N\,{\sc ii}]
$\lambda$5755 line (e.g. Crowther, Hillier \& Smith 1995). After
correction for instrumental width, the FWHMs of 4.81$\pm$0.27 \AA
\, (2015 June 14), 5.04$\pm$0.26 \AA \, (2015 August 2) and
5.05$\pm$0.20 (September 8) correspond to $v_\infty$ of 87$\pm$45,
117$\pm$33 and 118$\pm$21 $\kms$, respectively. Note that although
all five estimates agree with each other within the error margins,
the lower values of $v_\infty$ based on the [N\,{\sc ii}] line
might be because this line originates closer to the star where the
wind is still accelerated (Stahl et al. 1991). This supposition is
supported by the observed [N\,{\sc ii}] $\lambda\lambda$5755, 6584
line intensity ratio, which implies that the nitrogen forbidden
lines originate in a dense ($\geq$$10^7 \, {\rm cm}^{-3}$) matter
(e.g. Tarter 1969), probably close to the star (cf. Stahl et al.
1991). Note also that the low wind velocity of MN44 during its
minimum light phase suggests that this star is close to the
Eddington limit, probably because it already lost a significant
fraction of its initial mass either in a single giant eruption or
in a number of less catastrophic mass-loss episodes (Humphreys et
al. 2014a,b).

The forbidden lines can also be used to derive the systemic
velocity (Stahl et al. 2001). Inspection of
Table\,\ref{tab:inten}, however, shows significant changes in RVs
of the [Fe\,{\sc ii}] and [N\,{\sc ii}] lines from spectrum to
spectrum. One can see that the RVs of most of these lines
systematically reduced during the last four months and that the
velocity decrease was more prominent during the first month. RVs
of the Si\,{\sc ii} $\lambda\lambda$6347, 6371 absorption doublet
and the Balmer lines also demonstrate the same trend. The decrease
in the RVs of the Balmer lines would be much more stronger if one
takes into account the 2009's spectrum. Although these changes
suggest that MN44 might be a close eccentric binary system (cf.
Section\,\ref{sec:bin}), the limited available data did not allow
us to prove this for sure.

\begin{figure}
\begin{center}
\includegraphics[width=7cm,angle=270,clip=]{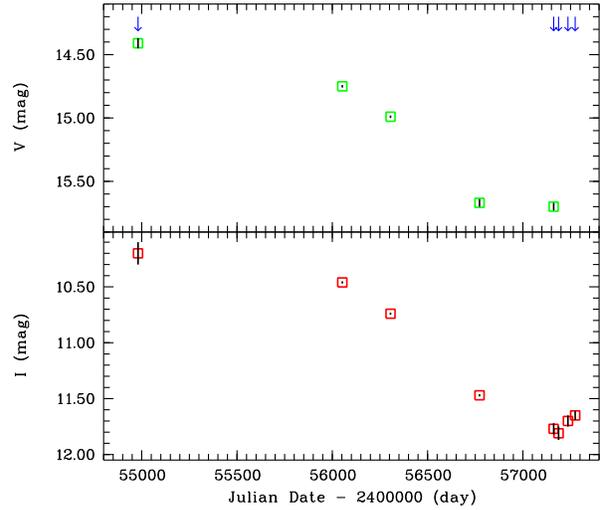}
\end{center}
\caption{Light curves of MN44 in the $V$ and $I$ bands. 1$\sigma$
error bars are indicated, but in most cases they are within the
size of the data points (boxes). The arrows mark the dates of the
spectroscopic observations.} \label{fig:phot}
\end{figure}

\begin{figure*}
\begin{center}
\includegraphics[angle=0,width=2\columnwidth,clip=]{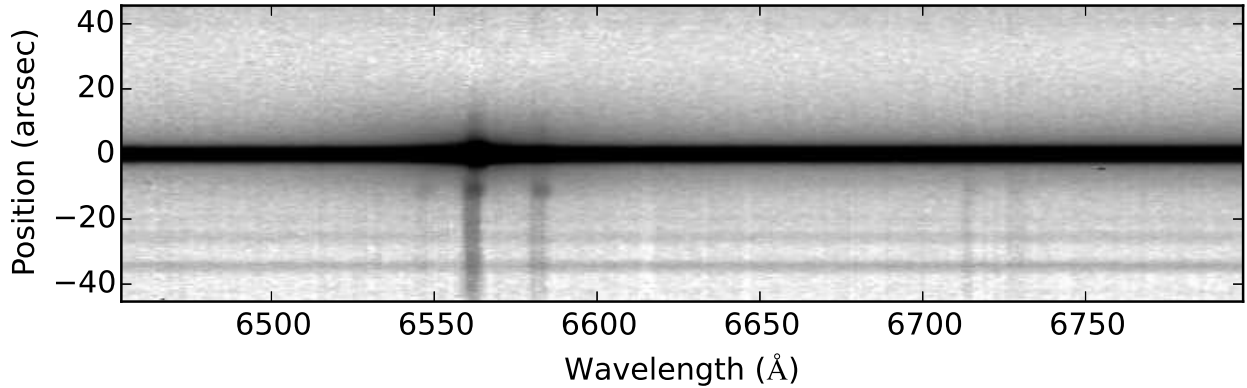}
\end{center}
\caption{A part of the 2D spectrum of the diffuse emission around
MN44 (obtained on 2015 June 14), showing nebular emission lines
(from left to right) of [N\,{\sc ii}] $\lambda$6548, H$\alpha$,
[N\,{\sc ii}] $\lambda$6584 and [S\,{\sc ii}]
$\lambda\lambda$6717, 6731. The lower part of the spectrum
corresponds to the emission northwest of MN44. The lines in this
direction have two distinct velocity components, namely, in the
inner region ($\la$14 arcsec from MN44) they are shifted to the
red end of the spectrum compared to the lines originated at larger
angular distances (outer region). The upper part of the spectrum
corresponds to the emission southeast of MN44, which extends from
the star for $\approx$10 arcsec. See Fig.\,\ref{fig:slit} and text
for details.} \label{fig:2D}.
\end{figure*}

The detected spectral variability of MN44 strongly suggests that
this star is a bona fide LBV, currently evolving towards the hot
state, i.e. towards the brightness minimum (e.g. Stahl et al.
2001). This implication is reinforced by photometric measurements
given in Table\,\ref{tab:phot}, which show that MN44 has
experienced strong brightness decline during the period of our
spectroscopic observations. This is illustrated by
Fig.\,\ref{fig:phot}, which plots $V$- and $I_{\rm c}$-band light
curves of MN44 since its identification as a cLBV in 2009 with
arrows indicating times when the spectra were obtained. For each
data point (square) we give 1$\sigma$ error bars, which in most
cases are less than the size of the squares themselves.
Fig.~\ref{fig:phot} shows that MN44 became fainter in the $V$ and
$I_{\rm c}$ bands, respectively, by $\approx$1.3 and 1.6 mag
during the last 6 years. One can see also that the $I_{\rm
c}$-band brightness of MN44 has reached minimum in 2015 June and
then starts to increase. On a longer time-scale, the $I_{\rm
c}$-band variability is even more striking. From
Table\,\ref{tab:phot} it follows that MN44 has brightened by
$\approx$3 mag (!) during $\approx$30 yr preceding our monitoring
campaign. The large magnitude of the $I_{\rm c}$-band excursion
suggests that MN44 might belong to a rare group of LBVs exhibiting
giant eruptions (as in $\eta$\,Car and P\,Cyg) during which their
bolometric luminosity changes (Humphreys \& Davidson 1994).
Table\,\ref{tab:phot} also shows that MN44 became brighter in the
$J$ and $K_{\rm s}$ bands by $\approx$0.6 mag during $\approx$7.5
months in 1998--1999.

By using the weighted least square approximation, we found that
the $V$$-$$I_{\rm c}$ colour of MN44 has decreased by $\approx$0.3
mag during the last six years. On the other hand, as one can infer
from Table\,\ref{tab:phot}, this decrease has occurred mostly
during the last year, while in 2009--2014 the colour did not
change much despite of significant ($>$1 mag) brightness decrease
of the star. Such behaviour is not typical of LBV excursions at
constant bolometric luminosity during which stars become bluer
with the brightness decrease, but it was observed for LBVs showing
changes in the bolometric luminosity (e.g. Clark et al. 2009;
Kniazev et al. 2015). Whether this is the case for MN44 as well
could be proved with a detailed spectral modelling, which is,
however, beyond the scope of this paper.

Taken together, our observations of MN44 unambiguously show that
this star is a new (seventeenth) member of the family of Galactic
bona fide LBVs (see Kniazev et al. 2015 for a recent census of
these objects).

\begin{figure}
\begin{center}
\includegraphics[angle=0,width=0.8\columnwidth,clip=]{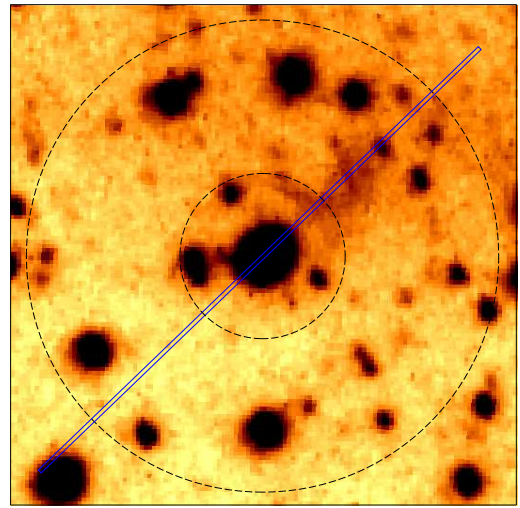}
\end{center}
\caption{SHS image of MN44 and the diffuse H$\alpha$+[N\,{\sc ii}]
emission around it with the orientation of the RSS slit
(PA=43$\degr$) shown by a (blue) rectangle. Concentric, dashed
circles of radius of 14 and 40 arcsec are overplotted on the image
to show the extraction regions of the 1D spectra of the diffuse
emission to the northwest of MN44. Note that the radius of the
inner circle is almost equal to the radius of the 24\,$\mu$m
circumstellar shell. See text for details.} \label{fig:slit}
\end{figure}

\begin{figure}
\begin{center}
\includegraphics[angle=270,width=1.0\columnwidth,clip=]{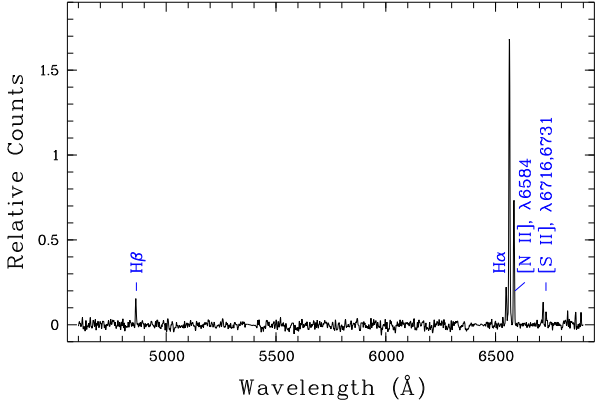}
\end{center}
\caption{1D spectrum of the knot to the northwest of MN44 obtained
on 2015 June 14.} \label{fig:Neb}
\end{figure}

\subsection{H$\alpha$+[N\,{\sc ii}] emission to the northwest of MN44}
\label{sec:ha}

The SHS H$\alpha$+[N\,{\sc ii}] image presented in
Fig.\,\ref{fig:neb} shows that there is a diffuse emission
immediately around MN44 and further northwest from it. There is
also a bright knot of emission apparently in contact with the
circumstellar shell. To clarify the nature of the emission around
MN44, we obtained long-slit SALT spectra with the slit oriented at
PA=43$\degr$ in order to cross the knot. In two-dimensional (2D)
spectra the knot appears as emission lines of H$\beta$, H$\alpha$,
[\ion{N}{ii}] $\lambda\lambda$6548, 6584 and [\ion{S}{ii}]
$\lambda\lambda$6717, 6731. In the following, we use the spectrum
taken on 2015 May 6 because of its better quality. A part of this
spectrum is presented in Fig.\,\ref{fig:2D}. One can see two
distinct velocity components in the emission lines. The lines
produced in the inner region ($\leq$14 arcsec from MN44, i.e.
within the confines of the 24\,$\mu$m shell) are shifted to the
red end of the spectrum compared to the lines originated at larger
angular distances from the star (outer region).

1D spectra of the inner and outer regions were extracted by
summing up, without any weighting, all rows in angular distance
intervals from $\approx$9 to 14 arcsec and from 14 to 40 arcsec,
respectively (see Fig.\,\ref{fig:slit}). The emission lines
detected in these spectra were measured using the programs
described in Kniazev et al. (2004). Table\,\ref{tab:int} lists the
observed intensities of the lines normalized to H$\beta$,
F($\lambda$)/F(H$\beta$), the reddening-corrected line intensity
ratios, I($\lambda$)/I(H$\beta$), the logarithmic extinction
coefficients, $C$(H$\beta$), the colour excesses, $E(B-V)$, and
the heliocentric radial velocities, $v_{\rm hel}$, of the two
regions. The 1D spectrum of the inner region is shown in
Fig.\,\ref{fig:Neb}.

\begin{table}
\centering{\caption{Line intensities in the spectra of diffuse
emission to the northwest of MN44.} \label{tab:int}
\begin{tabular}{lcc} \hline
$\lambda_{0}$(\AA) Ion &
F($\lambda$)/F(H$\beta$)&I($\lambda$)/I(H$\beta$) \\ \hline
& inner region & \\
\hline
4861\ H$\beta$\         &  1.00$\pm$0.10  &  1.00$\pm$0.10 \\
6548\ [N\ {\sc ii}]\    &  1.97$\pm$0.15  &  0.40$\pm$0.03 \\
6563\ H$\alpha$\        & 14.56$\pm$1.04  &  2.92$\pm$0.23 \\
6584\ [N\ {\sc ii}]\    &  5.69$\pm$0.41  &  1.12$\pm$0.09 \\
6717\ [S\ {\sc ii}]\    &  0.97$\pm$0.11  &  0.17$\pm$0.02 \\
6731\ [S\ {\sc ii}]\    &  0.49$\pm$0.08  &  0.09$\pm$0.02 \\
 & & \\
C(H$\beta$)\ dex        & \MC {2}{c}{2.10$\pm$0.09} \\
$E(B-V)$                & \MC {2}{c}{1.43$\pm$0.06 mag} \\
$v_{\rm hel}$           & \MC {2}{c}{$-$11$\pm$6 \, $\kms$} \\
\hline & outer region & \\ \hline
4861\ H$\beta$\            & 1.000$\pm$0.291 & 1.000$\pm$0.299 \\
6548\ [N\ {\sc ii}]\       & 0.745$\pm$0.234 & 0.196$\pm$0.066 \\
6563\ H$\alpha$\           &11.170$\pm$2.316 & 2.908$\pm$0.673 \\
6584\ [N\ {\sc ii}]\       & 2.269$\pm$0.504 & 0.583$\pm$0.144 \\
6717\ [S\ {\sc ii}]\       & 1.270$\pm$0.355 & 0.300$\pm$0.091 \\
6731\ [S\ {\sc ii}]\       & 1.028$\pm$0.316 & 0.241$\pm$0.080 \\
  & & \\
$C$(H$\beta$)                & \MC {2}{c}{1.76$\pm$0.27} \\
$E(B-V)$                     & \MC {2}{c}{1.20$\pm$0.18 mag} \\
$v_{\rm hel}$              & \MC {2}{c}{$-$39$\pm$12 \, $\kms$} \\
\hline
\end{tabular}
 }
\end{table}

The [S\,{\sc ii}] $\lambda\lambda$6716, 6731 line intensity ratio
can be used to derive the electron number density, $n_{\rm e}$, in
the line-emitting gas (Krueger, Aller \& Czyzak 1970; Saraph \&
Seaton 1970). In the inner region, the measured ratio of
$1.89^{+0.82} _{-0.53}$ agrees at 1 sigma level with the
theoretical upper limit on this ratio of $\approx 1.4$ (Krueger et
al. 1970). Thus, one can only put an upper limit on $n_{\rm e}$ of
$\leq 10 \, {\rm cm}^{-3}$. In the outer region, the line ratio of
$1.24^{+1.19} _{-0.59}$ corresponds to $n_{\rm e}=190^{+2770}
_{-180}$.

The above estimates of $C$(H$\beta$) and $n_{\rm e}$([S\,{\sc
ii}]) were derived for the Case\,B recombination and under the
assumption that the electron temperature, $T_{\rm e}$, is equal to
$10^4$ K.

The [N\,{\sc ii}] and [S\,{\sc ii}] line intensities can be used
to estimate the nitrogen to sulphur abundance ratio, $N({\rm N}^+
)/N({\rm S}^+ )$, which is almost independent of $n_{\rm e}$ and
$T_{\rm e}$, provided that $n_{\rm e}\leq 1000 \, {\rm cm}^{-3}$
and $T_{\rm e}\leq 10^4$ K. In this case, it is given by (see
Benvenuti, D'Odorico \& Peimbert 1973 and references therein):
\begin{equation}
{N({\rm N}^+ )\over N({\rm S}^+ )} =3.61 {I(6584) \over
I(6716+6731)} \, . \label{eq:abud}
\end{equation}
Using equation\,(\ref{eq:abud}) and Table\,\ref{tab:int}, one
finds $N({\rm N}^+ )/N({\rm S}^+ )$$\approx$15.55$^{+4.31}
_{-3.16}$ and 3.89$^{+3.20} _{-1.66}$ for the inner and outer
regions, respectively. The former ratio is $\approx$3.0$^{+0.9}
_{-0.6}$ times larger than the solar value of 5.12 (Asplund et al.
2009), while the latter one is comparable to it. The high $N({\rm
N}^+ )/N({\rm S}^+ )$ ratio in the inner region could be owing to
pollution by CNO-processed ejecta from MN44. Since the sulfur
abundance is not expected to change during stellar evolution, one
can argue that the inner region is enriched in N by a factor of
$\approx$2--4. This provides further support to the physical
association between the knot and the circumstellar shell around
MN44. The solar abundance of N in the outer region suggests that
it is an \hii region.

Since the $E(B-V)$ values for the inner and outer regions are
consistent with each other within the error margins, it is likely
that both are located at the same distance and ionized by MN44
during its hot state. If so, then the difference in the
heliocentric radial velocities of the two regions could be
understood if the inner part of the knot was accelerated because
of interaction with the circumstellar shell of MN44. This, in
turn, would imply that the radial velocity of the shell (and MN44)
relative to the \hii region is 28$\pm$$13 \, \kms$.

\subsection{Reddening to and luminosity of MN44}
\label{sec:lum}

The colour excess towards the knot could be used to derive the
$V$-band extinction, $A_V$, in the direction of MN44, and thereby
to constrain the absolute visual magnitude, $M_V$, of this star.
For the standard ratio of total to selective extinction of
$R_V$=3.1, one finds $A_{\rm V}=4.43$ mag, while for $M_V$ one has
the following equation:
\begin{equation}
M_V=V-{\rm DM}-A_V \, , \label{eq:MV}
\end{equation}
where DM is the distance modulus. Using equation\,(\ref{eq:MV})
and the $V$-band magnitude of MN44 of 15.69 (measured on 2015 May
19), one has $M_V$=11.26$-$DM mag. To further constrain $M_V$, we
note that the sightline towards MN44 first enters the
Carina-Sagittarius arm (located at a distance of $\sim$2.1 kpc
from the Sun), then (at $\sim$3.3 kpc) intersects the Crux-Scutum
arm, then twice crosses the Norma arm (at $\sim$5.5 and 10.8 kpc),
and then again crosses the Crux-Scutum and Carina-Sagittarius arms
(at $\sim$15 and 19 kpc, respectively) (e.g. Cordes \& Lazio
2002). The distances at which the sightline intersects the spiral
arms correspond to DM values of $\approx$11.61, 12.59, 13.70,
15.17, 15.88 and 16.39 mag, respectively. One can see that $M_V$
of MN44 would be at the upper end of values typical of B
supergiants even if this star is located in the more distant part
of the Carina-Sagittarius arm (DM=16.39 mag), i.e.
$M_V$$\approx$$-$5.1 mag. To derive the bolometric luminosity of
MN44, we note that the 2015's spectra of this star are very
similar to those of the LBV WS1 in the cool state, for which we
derived the effective temperature, $T_{\rm eff}$, of
$\approx$12\,000\,K (Kniazev et al. 2015). At this temperature,
the bolometric correction of MN44 is equal to $\approx$$-$0.8 mag.
Correspondingly, the bolometric luminosity of MN44 (for DM=16.39
mag) would be only of $\log(L_{\rm bol}/\lsun)$$\approx$4.3, which
is much smaller than the minimum value of 5.1 ever derived for
known bona fide or cLBVs (Humphreys et al. 2014a).

The reddening towards MN44 could also be estimated by matching the
dereddened spectral slope of this star with those of stars of
similar $T_{\rm eff}$. Using the Stellar Spectral Flux Library by
Pickles (1998), we found $E(B-V)$=3.30$\pm$0.10 mag (this estimate
only slightly depends on the assumed $T_{\rm eff}$; see Gvaramadze
et al. 2012a) and $A_V$=10.23 mag. Correspondingly, one has
$M_V$=5.46$-$DM mag. One can see that MN44 would has a reasonable
luminosity, i.e. $\log(L_{\rm bol}/\lsun)$$\geq$5.1, if the star
is located at $d$$\geq$3.3 kpc, i.e. in the near segment of the
Crux-Scutum arm or further out (DM$\geq$12.59 mag; see
Table\,\ref{tab:DM}). At these distances, MN44 would be located on
the cool side of the S\,Dor instability strip (Wolf 1989; Groh et
al. 2009) on the Hertzsprung-Russell diagram.

\begin{table*}
  \caption{MN44: absolute visual magnitude, luminosity, initial mass, lifetime, height below the Galactic
  plane, and peculiar velocity perpendicular to the Galactic plane for different distances/distance
  moduli and $A_V=10.23$ mag. See text for details.}
  \label{tab:DM}
  \renewcommand{\footnoterule}{}
  \begin{center}
  \begin{tabular}{cccccccc}
      \hline
      $d$ & DM & $M_V$ & $\log(L_{\rm bol}/\lsun)$ & $M_{\rm init}$ & $t$ & $h$ & $h/t$\\
      (kpc) & (mag) & (mag) &  & ($\msun$) & (Myr) & (pc) & ($\kms$) \\
      \hline
      2.1 & 11.61 & $-$6.2 & 4.7 & 15 & 12 & 40 & 3 \\
      3.3 & 12.59 & $-$7.1 & 5.1 & 20 & 9 & 70 & 8 \\
      5.5 & 13.70 & $-$8.2 & 5.5 & 35 & 6 & 110 & 18 \\
      10.8 & 15.17 & $-$9.7 & 6.1 & 60 & 4 & 220 & 55 \\
      15 & 15.88 & $-$10.4 & 6.4 & 120 & 3 & 300 & 100 \\
      19 & 16.39 & $-$10.9 & 6.6 & 150 & 2.5 & 380 & 150 \\
      \hline
    \end{tabular}
    \end{center}
    \end{table*}

The $A_V$ value derived from the spectral slope of MN44 exceeds by
$\approx$5.8 mag the extinction derived towards the knot. A
trivial explanation of this discrepancy is that MN44 and the knot
are located at different distances and are simply projected by
chance along the same line-of-sight. This explanation, however, is
unlikely because there are strong indications that the knot is in
a contact with the circumstellar shell of MN44. Instead, one can
envisage two other possible explanations.

Our preferred explanation of the large discrepancy in the $A_V$
values is that MN44 is shrouded by dense circumstellar material
and that this material significantly contributes to the extinction
of the star. For example, the dusty material of the Homunculus
nebula is responsible for $\approx$4 mag of the $V$-band
extinction towards $\eta$\,Car (Humphreys, Davidson \& Smith 1999;
cf. van Genderen, de Groot \& The 1994). It is, therefore,
possible that the dusty ejecta from recent outbursts of MN44
causes the excess extinction towards this star as well. If this
explanation is correct, then one can expect that the changes in
the brightness of MN44 might be caused not only by the intrinsic
variability of the star itself, but also by the changeable
circumstellar extinction (cf. van Genderen et al. 1994). Or else
the extra extinction might be due to a dense circumbinary disk if
MN44 is a binary system (see Section\,\ref{sec:bin}).

An alternative explanation of the discrepancy is that the
intrinsic Balmer decrement in the spectrum of the knot is much
shallower than what follows from the standard Case\,B
recombination model, i.e. the intrinsic value of the
$I$(H$\alpha$)/$I$(H$\beta$) ratio is significantly smaller than
$\approx$3. For this explanation to work, the H emission should
arise in a very dense medium with $n_{\rm e}\ga10^{13} \, {\rm
cm}^{-3}$ (Drake \& Ulrich 1980). This requirement does not
contradict to the low electron density indicated by the [S\,{\sc
ii}] $\lambda\lambda$6716, 6731 line intensity ratio because these
forbidden lines could originate in a low-density halo around the
dense core of the knot. If the H emission indeed originates in the
very dense medium, then the $I$(H$\alpha$)/$I$(H$\beta$) ratio
could be as small as $\approx$1 (see fig.\,7 in Drake \& Ulrich
1980), which for the line intensities observed in the spectrum of
the knot would result in a factor of $\approx$2 larger values of
$E(B-V)$ and $A_{\rm V}$, thereby bringing them closer to the
values derived from the dereddening the spectral slope of MN44. To
check this explanation, one has to measure more Balmer line ratios
in the spectrum of the knot, which requires much more deep
spectroscopic observations.

An additional constraint on the distance to MN44 comes from the
empirical amplitude-luminosity relation of LBVs, which implies
that the amplitude of photometric variability increases with
$L_{\rm bol}$ (Wolf 1989). The strong changes in the brightness of
MN44 suggest that the luminosity of this star should be well above
the lower end of the range of luminosities derived for (c)LBVs.
This means that the star should be located at least in the Norma
arm or in the next arms out, i.e. at $d$$\geq$5.5 kpc. At these
distances, MN44 would lie at $\geq$110 pc below the Galactic
plane. This separation from the plane is comparable to or larger
than the exponential scale height of runaway O stars of
$\approx$90 pc (Stone 1979) and is at least a factor of two larger
than that of ``normal" OB stars (Stone 1979; Reed 2000). Thus,
MN44 might be a runaway star; we further discuss this possibility
in Section\,\ref{sec:run}.

On the other hand, the larger the distance to MN44, the larger its
luminosity and initial mass, $M_{\rm init}$, the shorter the
lifetime of this star, $t$, and the larger its separation from the
Galactic plane, $h=d\sin b$ (see Table\,\ref{tab:DM}). This means
that the more massive the star the higher the peculiar velocity
perpendicular to the Galactic plane, $\sim$$h/t$, it should have
(see Table\,\ref{tab:DM}). For example, if MN44 is located at
$d$=15 kpc ($M_{\rm init}\approx$$120 \, \msun$) then its short
lifetime of $t\sim$3 Myr would imply that to reach the height of
$h\approx$300 pc, the star should be ejected from the parent
cluster with a velocity of $\approx$$100 \, \kms$ (provided that
the ejection event has occurred soon after the star was born).
Although ejection of such massive and high-velocity stars is not
impossible, the probability of these events is very low
(Gvaramadze \& Gualandris 2011). It would be even much lower if
MN44 is a massive binary system (see next section).

Finally, the distance to MN44 could also be constrained through
the kinematic distance to the \hii region northwest of the star
(in Sect.\,\ref{sec:ha}, we argue that they are likely located at
the same distance). Using the heliocentric radial velocity of the
\hii region of $-$39$\pm$$12 \, \kms$, adopting the radial
velocity dispersion of \hii regions within an arm of $\sim$$10 \,
\kms$ (Georgelin \& Georgelin 1976), and assuming the distance to
the Galactic Centre of $R_0$=8.0 kpc and the circular rotation
speed of the Galaxy of $\Theta _0 =240 \, \kms$ (Reid et al.
2009), and the solar peculiar motion
$(U_{\odot},V_{\odot},W_{\odot})=(11.1,12.2,7.3) \, \kms$
(Sch\"onrich, Binney \& Dehnen 2010), one finds a kinematic
distance of $\approx$$2.3^{+1.1} _{-1.4}$ or $12.1^{+1.4} _{-1.1}$
kpc. For the distance range implied by the error margins of the
former distance estimate, the luminosity of MN44 would be
unrealistically low (see above). The error margins of the latter
distance estimate allow the possibility that MN44 is located in
the outer segment of the Norma arm. In this case, the bolometric
luminosity of MN44 of $\log(L_{\rm bol}/\lsun)$=6.1 would exceed
the Humphreys-Davidson luminosity limit (Humphreys \& Davidson
1979), i.e. the star would be located in a region of the
Hertzsprung-Russell diagram where the massive stars are expected
to experience unsteady (eruptive) high mass-loss episodes. We
caution however, that the possible association between the
optically visible knot and the circumstellar shell is rather
indicative of moderate ($\approx$4 mag) interstellar extinction
towards MN44, which in turn suggests that this star is located at
a shorter distance, i.e. in the inner segment of the Norma arm. In
fact, the above kinematic distance estimates would be meaningless
if the \hii region has a peculiar radial velocity of several tens
of $\kms$, e.g. because of interaction with the stellar
wind/ejecta during the previous mass-loss episodes.

\subsection{MN44: a massive X-ray binary?}
\label{sec:bin}

In Section\,\ref{sec:neb}, we noted that Masetti et al. (2010)
identified MN44 as a counterpart of the {\it INTEGRAL} source of
hard X-ray emission IGR\,J16327$-$4940 and tentatively classified
this source as a HMXB. Let us discuss whether the LBV
classification of MN44 is consistent with the possibility that
this object is a source of hard X-ray emission.

First, we list some relevant data on IGR\,J16327$-$4940 given in
the Fourth soft gamma-ray source catalogue obtained with the IBIS
gamma-ray imager on board the {\it INTEGRAL} satellite (Bird et
al. 2010). In this catalogue, IGR\,J16327$-$4940 is indicated as a
transient (strongly variable) source with the 20--40 and 40--100
keV band fluxes time-averaged over the total exposure time of
3319.3 ks (or $\approx$38.4 d) of $F$(20--40)$<$0.2 mCrab and
$F$(40--100)=0.4$\pm$0.1 mCrab, respectively. The catalogue also
gives the peak 20--40 keV band flux of $F_{\rm
peak}$(20--40)=2.1$\pm$0.7 mCrab, i.e. the mean flux measured
during the largest observed outburst, in which the source was
detected at a maximum level of significance of 5.6-$\sigma$.

The above fluxes translate to the X-ray-to-bolometric luminosity
ratios of $\log[L_{\rm X}(20-40)/L_{\rm bol}]<-5.30$, $\log[L_{\rm
X}(40-100)/L_{\rm bol}]=-4.91$ and $\log[L_{\rm X}(20-40)/L_{\rm
bol}]=-4.28$, which are independent of the actual distance to
MN44. These ratios should be compared with $\log[L_{\rm X}/L_{\rm
bol}]$ values observed for other (c)LBVs.

Naz\'{e}, Rauw \& Hutsem\'{e}kers (2012) conducted a search for
X-ray emission from the Galactic bona fide and cLBVs using
available at that time data from the {\it XMM-Newton} and {\it
Chandra} X-ray observatories, which covered 31 of 67 known objects
of this type. Besides two already know X-ray sources $\eta$\,Car
and NAME\,VI\,CYG\,12, they reported detection of X-ray emission
from two additional (c)LBVs, Cl*\,Westerlund\,1\,W\,243 and
GAL\,026.47+00.02, as well as possible detections of two more
(c)LBVs near the Galactic Centre, GCIRS\,34W and GCIRS\,33SE. For
four (c)LBVs with confidently detected X-ray emission the
following $\log[L_{\rm X}/L_{\rm bol}]$ values were derived:
$\sim$$-$5 ($\eta$\,Car), $-$6.1 (NAME\,VI\,CYG\,12), $-$5.95
(GAL\,026.47+00.02) and $-$7.0...7.3 (Cl*\,Westerlund\,1\,W\,243).
These values imply that the $L_{\rm X}/L_{\rm bol}$ ratios of the
first three stars are one or two orders of magnitude larger than
those typical of single O stars (i.e. $\approx$$10^{-7}$; e.g.
Pallavicini et al. 1981; Sana et al. 2006), whose X-ray emission
is believed to originate because of shocks inside stellar winds
(e.g. Owocki, Castor \& Rybicki 1988; Feldmeier, Puls \& Pauldrach
1997). Since the wind velocities of the majority of (c)LBVs
($\sim$$100 \, \kms$) are about an order of magnitude lower than
those of O stars, the bright X-ray emission of the three stars
cannot be caused by intrinsic wind shocks, but rather should
originate in colliding winds (Usov 1992; Stevens, Blondin \&
Pollock 1992), i.e. these stars should be members of close massive
binaries (Naz\'{e} et al. 2012). This assertion conforms with the
observed hardness and variability of the X-ray emission from two
of these stars -- $\eta$\,Car and NAME\,VI\,CYG\,12.
Cl*\,Westerlund\,1\,W\,243 is also likely to be a massive binary
system because its slow wind cannot produce wind shocks strong
enough to account for the observed X-ray emission. If so, then the
X-ray emission from this object should originate in a companion O
star (Naz\'{e} et al. 2012).

For the remaining 23 (c)LBVs observed with {\it Chandra} and {\it
XMM-Newton}, only upper limits on their X-ray fluxes were
obtained. In most cases, these limits imply that $\log[L_{\rm
X}/L_{\rm bol}]<-5...6$, which leaves the possibility that the
corresponding stars might be X-ray bright close massive binaries.
For five (c)LBVs very strong constraints on the X-ray emission
were derived, $\log[L_{\rm X}/L_{\rm bol}]<-8.2...9.4$, indicating
that some (c)LBVs are intrinsically weak X-ray sources. Possible
reasons for this are that these (c)LBVs are single stars and/or
their X-ray emission is significantly attenuated by dense
circumstellar material (Naz\'{e} et al. 2012).

The high values of $\log[L_{\rm X}/L_{\rm bol}]$ of MN44 along
with the hardness and variability of its X-ray emission point to
the possibility that this star is a colliding-wind binary. In this
case, the observed X-ray emission should originate at the shock
interface between colliding winds, while its variability could be
due to changes in the strength of the shocks caused by the
eccentricity of the binary orbit and/or the S\,Dor-like
variability of wind parameters (velocity, mass-loss rate) of the
LBV star. Interestingly, at the distance of 5.5 kpc, the 20--100
keV band X-ray luminosity of MN44 of 1.4$\times 10^{34} \, {\rm
erg} \, {\rm s}^{-1}$ would be comparable to that of $\eta$\,Car
of 0.7$\times$$10^{34} \, {\rm erg} \, {\rm s}^{-1}$ (Leyder,
Walter \& Rauw 2008). In this energy range, the X-ray emission
originates mostly from non-thermal processes. In the case of
$\eta$\,Car, it is believed that its 20--100 keV band X-ray
emission is dominated by inverse Compton scattering of stellar UV
photons by electrons accelerated in the wind collision zone
(Leyder et al. 2008). The same mechanism could be at work in MN44
as well.

If subsequent (optical or infrared) spectroscopic monitoring will
prove that MN44 forms a binary system with another massive star,
then dedicated X-ray observations of this system would be
desirable to search for orbital modulation of its X-ray emission
(e.g. Naz\'{e} et al. 2007; Hamaguchi et al. 2014). The X-ray
light curve, however, could not necessarily be periodic because
the orbital modulation of the X-ray emission might be affected by
changes in the wind of the LBV component of the binary (unless the
S\,Dor-like activity of MN44 is triggered by the effect of the
companion star). Also, an analysis of the {\it INTEGRAL} data
would be of interest to check if the changes in the hard X-ray
flux of MN44 are related to any particular orbital phase.

Alternatively, as suggested by Masetti et al. (2010), MN44 could
be a HMXB, i.e. a binary system composed of a massive star and a
compact object, either a neutron star or black hole (van den
Heuvel \& Heise 1972; Tutukov \& Yungelson 1973). In these
systems, the hard X-ray emission is generated because of accretion
of the stellar wind material onto the compact object, while its
variability could be due to changeable accretion rate. In the case
of MN44 the accretion rate could vary because of changes in the
wind velocity and mass-loss rate of the LBV companion star and/or
because of the eccentricity of the binary orbit. If the HMXB
nature of MN44 would be confirmed (e.g. by detection of pulsed
X-ray emission), then this system would represent a first known
example of a HMXB with a bona fide LBV donor star (cf. Mason et
al. 2012; Clark et al. 2013).

\subsection{MN44: a runaway star?}
\label{sec:run}

MN44, like the majority of other bona fide and cLBVs, is located
in the field and, therefore, is most probably a runaway star
(Gvaramadze et al. 2012a,b). The runaway status of MN44 is also
suggested by separation of this star from the Galactic plane (see
Sect.\,\ref{sec:lum}). To check this possibility, we searched for
proper motion measurements for MN44 using the VizieR catalogue
access tool\footnote{http://webviz.u-strasbg.fr/viz-bin/VizieR}.
We found four catalogues that provide proper motions for MN44,
namely, UCAC2 (Zacharias et al. 2004), PPMXL (R\"{o}ser,
Demleitner \& Schilbach 2010), SPM\,4.0 (Girard et al. 2011) and
UCAC4 (Zacharias et al. 2013). Since the measurement uncertainties
in the first two catalogues are larger than the measurements
themselves, we will use only the later two catalogues, which give
$\mu _\alpha \cos \delta =-7.86\pm2.25, \, \mu _\delta
=-2.59\pm2.41$ and $\mu _\alpha \cos \delta =-7.5\pm3.0, \, \mu
_\delta =-5.6\pm3.0$, respectively.

Using the same Galactic constants and the solar peculiar motion as
in Section\,\ref{sec:lum}, we calculated the peculiar transverse
velocity of MN44, $v_{\rm tr}=(v_{\rm l} ^2+v_{\rm b} ^2)^{1/2}$,
where $v_{\rm l}$ and $v_{\rm b}$ are the velocity components in
the Galactic coordinates. For the sake of illustration, we adopted
two plausible distances: 5.5 and 10.8 kpc. For the error
calculation, only the errors of the proper motion measurements
were considered. The results are summarized in
Table\,\ref{tab:prop}.

Taken at face value, the obtained peculiar velocities imply that
MN44 is a runaway star moving towards the Galactic plane, i.e. in
the ``incorrect" direction (recall that MN44 is located below the
Galactic plane). However, the large error margins of the velocity
components allow the possibility that MN44 is moving away from the
Galactic plane at 2$\sigma$ and 1$\sigma$ significance level for
the SPM\,4.0 and UCAC4 proper motion measurements, respectively.
Note also that the derived $v_{\rm tr}$ differs from zero only at
$\approx$2$\sigma$ or lower significance. Thus, to unambiguously
prove the runaway status of MN44, one needs more precise proper
motion measurements, which could be achieved with the space
astrometry mission Gaia.

\begin{table}
\caption{Peculiar transverse velocity and its components (in
Galactic coordinates) for two adopted proper motion measurements
and distances.} \label{tab:prop}
\begin{center}
\begin{minipage}{\textwidth}
\begin{tabular}{ccccc}
\hline
Sources for & $d$ & $v_l$ & $v_b$ & $v_{\rm tr}$ \\
proper motions & kpc & ($\kms$) & ($\kms$) & ($\kms$) \\
\hline
SPM\,4.0 & 5.5 & $-67$$\pm$61 & 99$\pm$61 & 120$\pm$61 \\
UCAC4 & 5.5 & $-118$$\pm$78 & 38$\pm$78 & 124$\pm$78 \\
SPM\,4.0 & 10.8 & 36$\pm$120 & 198$\pm$120 & 201$\pm$120 \\
UCAC4 & 10.8 & $-64$$\pm$154 & 79$\pm$154 & 102$\pm$154 \\
\hline
\end{tabular}
\end{minipage}
\end{center}
\end{table}

\section{Summary and conclusion}
\label{sec:con}

In this paper, we reported the discovery of a new (seventeenth)
Galactic bona fide LBV. The discovery was made through the
detection of a circular mid-infrared shell (of diameter of
$\approx$30 arcsec) with the {\it Spitzer Space Telescope} and
follow-up spectroscopic and photometric observations of its
central star -- MN44. The first epoch (2009) optical spectroscopy
of MN44 revealed a rich emission spectrum, typical of LBVs near
the visual maximum. The spectra taken six yr later showed the
emergence of He\,{\sc i} emission lines, which indicates that the
star became hotter ($T_{\rm eff}$$\approx$12\,000 K). Besides, the
EWs of the Balmer lines increased by $\approx$3--9 times, while
the forbidden lines of [N\,{\sc ii}] and [Fe\,{\sc ii}], and some
Fe\,{\sc ii} lines became much more prominent. Using the
flat-topped [Fe\,{\sc ii}] $\lambda$5376 line, we derived the
terminal wind velocity of MN44 of $\approx$$170 \, \kms$ during
the current hot state of this star, which could be considered as
an indication that MN44 is close to its Eddington limit.

Our photometric observations showed that the spectral variability
of MN44 is accompanied by the brightness decrease ($\sim$2 mag
during the last six yr), which is typical of LBVs recovering from
S\,Dor-type outbursts. From archival photometric data, we also
derived that MN44 has brightened in the $I_{\rm c}$ band by
$\approx$3 mag in $\approx$30 yr preceding our observations. The
large amplitude of photometric variability and the separation of
MN44 from the Galactic plane were used to suggest that this star
is located in the Norma arm, i.e. at $\sim$5.5 or 10.8 kpc. At
these distances, MN44 lies, respectively, $\approx$110 and 220 pc
below the Galactic plane, which implies that MN44 is a runaway
star.

Comparison of the {\it Spitzer} and SHS images revealed a knot of
H$\alpha$+[N\,{\sc ii}] emission to the northwest of MN44. The
knot partially delineates the mid-infrared circumstellar shell,
which is brighter in the place of possible contact with the knot.
We suggested that the shell is interacting with a density
inhomogeneity (knot) in the ambient medium. This suggestion was
reinforced by the detection of enhanced nitrogen abundance in the
knot, which might be caused by pollution from CNO-processed ejecta
from MN44. Using the Balmer decrement in the spectrum of the knot,
we derived the $V$-band extinction towards MN44 of $4.43$ mag.
This value turns out to be smaller by $\approx$6 mag than the
extinction derived from dereddening the spectral slope of MN44.
Two possible explanations of this discrepancy were discussed. The
first one (our preference) is that MN44 is shrouded by dusty
circumstellar material, which is responsible for most of the
reddening of the star. Alternatively, the intrinsic Balmer
decrement in the spectrum of the knot might be much shallower than
what follows from the Case\,B recombination model. This could
happen if the number density of the knot is $\ga$$10^{13} \, {\rm
cm}^{-3}$.

We discussed possible association of MN44 with the {\it INTEGRAL}
transient source of hard X-ray emission IGR\,J16327$-$4940. If
real, this association would imply that MN44 is either a
colliding-wind binary (similar to $\eta$\,Car and several other
Galactic (c)LBVs) or a first known example of a HMXB with a bona
fide LBV donor star.

To conclude, further spectroscopic and photometric monitoring of
MN44 with better resolution and cadence would be of great
importance for understanding the driving mechanisms of variability
of this star, and, potentially, for revealing its duplicity. If
the binary status of MN44 will be proved, then dedicated X-ray
observations of this system would be highly desirable to search
for orbital modulation of its X-ray flux (if the system is a
wind-colliding binary) or X-ray pulsations from the compact object
(if MN44 is a HMXB).

\section{Acknowledgements}
We are grateful to R.M.\,Humphreys (the referee) for useful
comments and suggestions on the manuscript. Some observations
reported in this paper were obtained with the Southern African
Large Telescope (SALT). VVG and LNB acknowledge the Russian
Science Foundation grants 14-12-01096 and 14-22-00041,
respectively. AYK acknowledges support from the National Research
Foundation (NRF) of South Africa. We thank Dave Kilkenny for
getting for us $V$- and $I_{\rm c}$-band images of MN44 with the
SAAO 1-m telescope. This work is based in part on archival data
obtained with the {\it Spitzer} Space Telescope, which is operated
by the Jet Propulsion Laboratory, California Institute of
Technology under a contract with NASA, and has made use of the
NASA/IPAC Infrared Science Archive, which is operated by the Jet
Propulsion Laboratory, California Institute of Technology, under
contract with the National Aeronautics and Space Administration,
the SIMBAD database and the VizieR catalogue access tool, both
operated at CDS, Strasbourg, France.

\end{document}